\documentclass{article}
\usepackage{mrs2005,epsfig,amssymb}
\def\na{{$N_{810}$}\/}
\def\nb{{$N_{817}$}\/}
\def\nc{{$N_{824}$}\/}
\def\med{{$M_{815}$}\/}
\def\lya{{Ly$\alpha$}\/}
\def\ha{{H$\alpha$}\/} 
\def\hb{{H$\beta$}\/} 
\def\oii{{[{{\sc O}\,{\sc ii}}]\/}} 
\def\oiii{{[{{\sc O}\,{\sc iii}}]\/}} 
\def\Oii{{[{{\sc O}\,{\sc ii}}]~$\lambda\lambda$3726,3728}\/} 

\def\ergsPerAng{erg\,s$^{-1}$\,cm$^{-2}$\,\AA$^{-1}$}
\def\ergscm2{erg\,s$^{-1}$\,cm$^{-2}$}
\def\ergs{erg\,s$^{-1}$}
\def\Mpc3{Mpc$^{3}$}
\def\Msunyr{M$_\odot$\,yr$^{-1}$}
\def\kms{km\,s$^{-1}$}

\begin{document} 
\providecommand{\ion}[2]{{\sc #1}\,{\sc #2}}
\title{WFILAS: WIDE FIELD IMAGER LYMAN ALPHA SEARCH} 
 
\author{EDUARD WESTRA$^1$, D.~Heath Jones$^1$, Chris Lidman$^2$,
  Ramana Athreya$^3$, Klaus Meisenheimer$^4$, Christian Wolf$^5$,
  Thomas Szeifert$^2$, Emanuela Pompei$^2$, Leonardo Vanzi$^2$}

\affil{$^1$Mt. Stromlo Observatory, Research School of Astronomy \&
  Astrophysics, Australia}
\affil{$^2$European Southern Observatory, Chile}
\affil{$^3$National Centre for Radio Astrophysics, India}
\affil{$^4$Max Planck Institute f\"ur Astronomie, Germany}
\affil{$^5$Department of Physics, University of Oxford, United
  Kingdom}
 
\begin{abstract} 
  The Wide Field Imager Lyman-Alpha Search (WFILAS) is a search for
  \lya\ emitting galaxies at $z\sim5.7$. Deep images from the Wide
  Field Imager (WFI) on the ESO/MPI 2.2\,m telescope have been used to
  detect 7 bright \lya\ emitting candidates in three fields covering
  0.74 sq.\,degree on the sky. For this we used three narrowband (FWHM
  $\sim$\,70\,\AA), one encompassing intermediate band (FWHM
  $\sim$\,220\AA) and broadband $B$ and $R$ filters. One has thus far
  been spectroscopically confirmed as a \lya\ emitting galaxy at
  $z=5.721$ using FORS2 at the VLT. This galaxy shows a bright, well
  resolved asymmetric line profile, which is characteristic of \lya\
  emitting galaxies.

  In one of our three fields, the Chandra Deep Field South (CDFS), we
  find an overdensity of \lya\ emitters in agreement with other
  surveys that have targeted this region. A statistically complete
  sample of our candidates probes the bright-end of the luminosity
  function, confirming earlier results from other smaller, deeper
  surveys.
\end{abstract} 

\section{Introduction} 
\label{sec:introduction}

To be able to derive the star formation history of the Universe, it is
necessary to determine the star formation rate at several redshifts.
Emission lines are good probes for this, as the star formation rate is
proportional to the line luminosity. The hydrogen Balmer series is
particularly well suited for this purpose. The Balmer emission line
luminosity is directly proportional to the total ionising flux of the
young OB stars embedded in \ion{H}{ii} regions, which is equatable to
the rate at which they form \cite{Kewley04}. Other lines, such as
\oii\ and \oiii, can be used as well, although the metallicity of the
surrounding gas needs to be taken into account. In all cases, the line
luminosities need to be corrected for internal extinction due to dust.
(\cite{Kewley04}; and references therein). It is also possible to use
\lya\ as star formation indicator, though extinction corrections are
much higher, because the line is in the rest-frame ultraviolet. By
using narrowband filter surveys it is possible to probe several of
these emission lines at different redshifts.

\section{WFILAS}
\label{sec:wfilas}
The Wide Field Imager Lyman Alpha Search (WFILAS) is a narrowband
filter survey designed to detect \lya\ emitting galaxies at
$z\sim5.7$. It was undertaken with the 2.2\,m ESO/MPI telescope at
Cerro La Silla with the Wide Field Imager (WFI). We targeted three
fields to be able to assess the influence of cosmic variance. The
fields are the Chandra Deep Field South (CDFS), a field near the South
Galactic Pole (SGP) and the COMBO-17 S11 field. This yields a total
sky coverage of 0.74 sq.\,degree, which gives a volume of
$1.0\times10^6$\,\Mpc3 at $z\sim5.7$\footnote{Throughout this paper we
  assume $H_0$\,=\,70\,km\,s$^{-1}$\,Mpc$^{-1}$, $\Omega_M$\,=\,0.3
  and $\Omega_\Lambda$\,=\,0.7}. Together with the relatively bright
survey limit ($m_{AB}\sim$\,24.0$-$24.5) we expect to find \lya\
emitting galaxies at the bright end of the luminosity function.

\begin{figure}[!htbp]
  \begin{center}
    \epsfig{figure=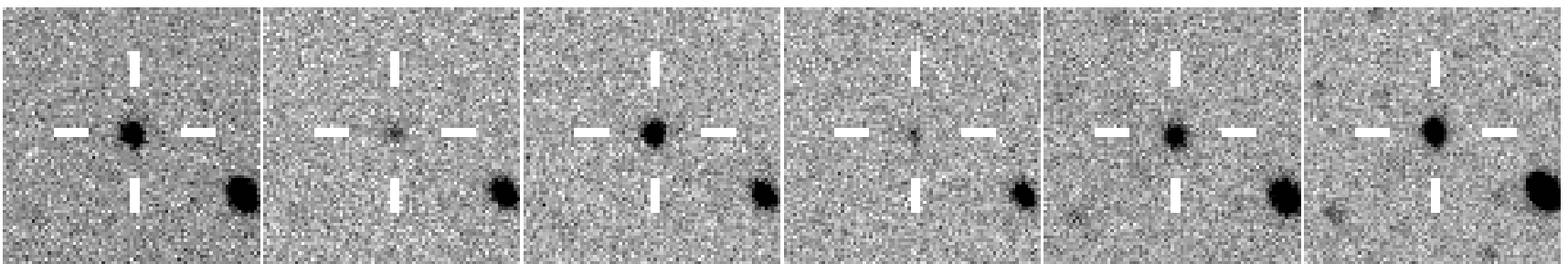,width=0.75\textwidth}
    \epsfig{figure=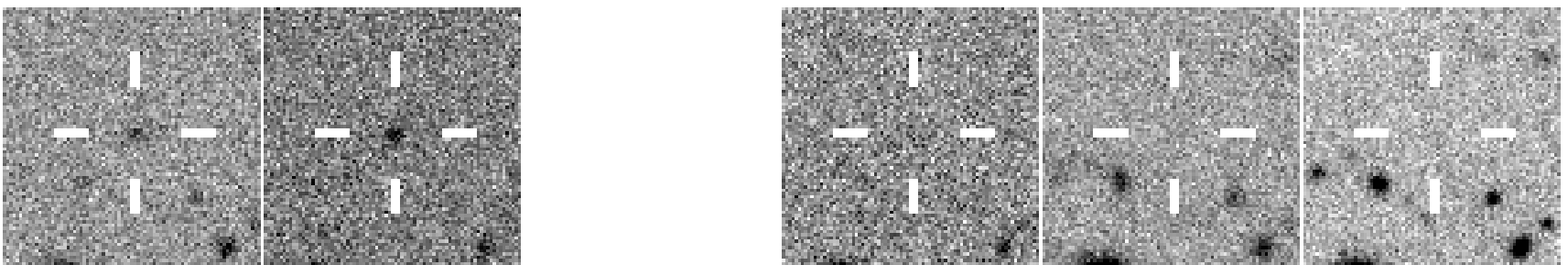,width=0.75\textwidth}
  \end{center}
\vspace{-0.4cm}
  \makebox[0pt][l]{\hspace{0.16\columnwidth} \med}
  \makebox[0pt][l]{\hspace{0.28\columnwidth} \na}
  \makebox[0pt][l]{\hspace{0.40\columnwidth} \nb}
  \makebox[0pt][l]{\hspace{0.515\columnwidth} \nc}
  \makebox[0pt][l]{\hspace{0.645\columnwidth} $B$}
  \makebox[0pt][l]{\hspace{0.764\columnwidth} $R$}
  \caption{Two examples of candidate line emitting galaxies. These
    figures demonstrate the contrast of the narrowband filters and the
    use of the broadband $B$ and $R$ data to distinguish between low
    and high redshift galaxies. From left to right are the
    intermediate band filter \med, three narrowband filters \na, \nb\
    and \nc, and the broadband filters $B$ and $R$. ({\it top}) Line
    emitting galaxy confirmed to be at $z=0.25$. The galaxy is clearly
    visible in \nb. The strong \ha\ line of this galaxy resides in the
    spectral range of this filter. In the neighbouring \na\ and \nc\
    filter the continuum of the galaxy can be seen. Furthermore, this
    galaxy is visible in both broadband filters, which confirm the low
    redshift nature of the galaxy. ({\it bottom}) Candidate $z\sim5.7$
    line emitting galaxy. In this case the possible \lya\ line resides
    in the spectral range of the \na\ filter. There is no \nb\
    available for the second object. Moreover, there is no detection
    in both broadband filters, strengthening the case of the line
    being \lya. If confirmed, this galaxy will have the strongest
    \lya\ line at this redshift
    ($\sim$\,1$\times$10$^{-16}$\,\ergscm2).}
  \label{fig:examples}
\end{figure}

This survey is unique in its filter setup. Instead of the usual single
narrowband filters we have used three neighbouring narrowband filters
encompassed by one intermediate band filter (FWHM\,$\sim$\,220\,\AA).
The particular narrow bandpass (FWHM\,$\sim$\,70\,\AA) of the
narrowband filters provides a high contrast for line emitting
galaxies. This is demonstrated in Figure \ref{fig:examples}. Two
examples are given, a low redshift galaxy (spectrally confirmed to be
an \ha\ emitting galaxy at $z=0.25$) and a candidate high redshift
galaxy (still to be confirmed). In both cases the galaxy is visible in
the narrowband image in which emission line resides, particulary in
the low redshift case.

Using the images that were taken in the broadband $B$ and $R$
filters\footnote{The broadband $B$ and $R$ and a fraction of the
  intermediate band \med\ data have been taken from the COMBO-17
  survey \cite{Wolf03}}, we are able to minimise the contamination of
low redshift line emitting galaxies, as \lya\ galaxies have no
continuum bluewards of the \lya\ line. At the same time, we can use
these low redshift candidates to measure the star formation rate at
their respective redshifts. We will touch on this aspect of the survey
in Section \ref{sec:ttffgs} in combination with the Taurus Tuneable
Filter Field Galaxy Survey.

\begin{figure}[!htbp]
  \begin{center}
    \epsfig{figure=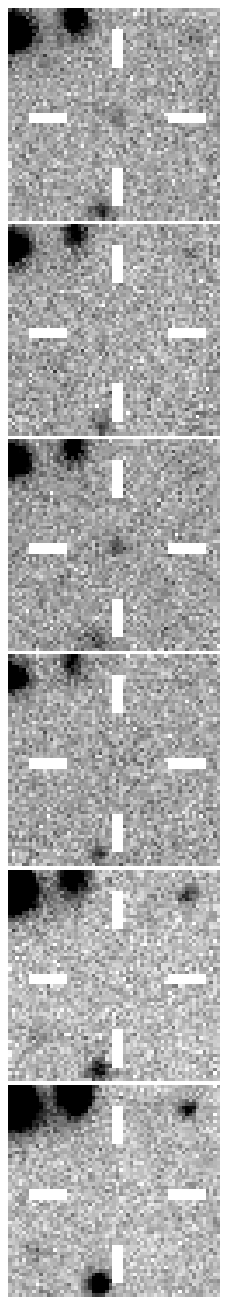,height=0.66\textheight,clip}
    \epsfig{figure=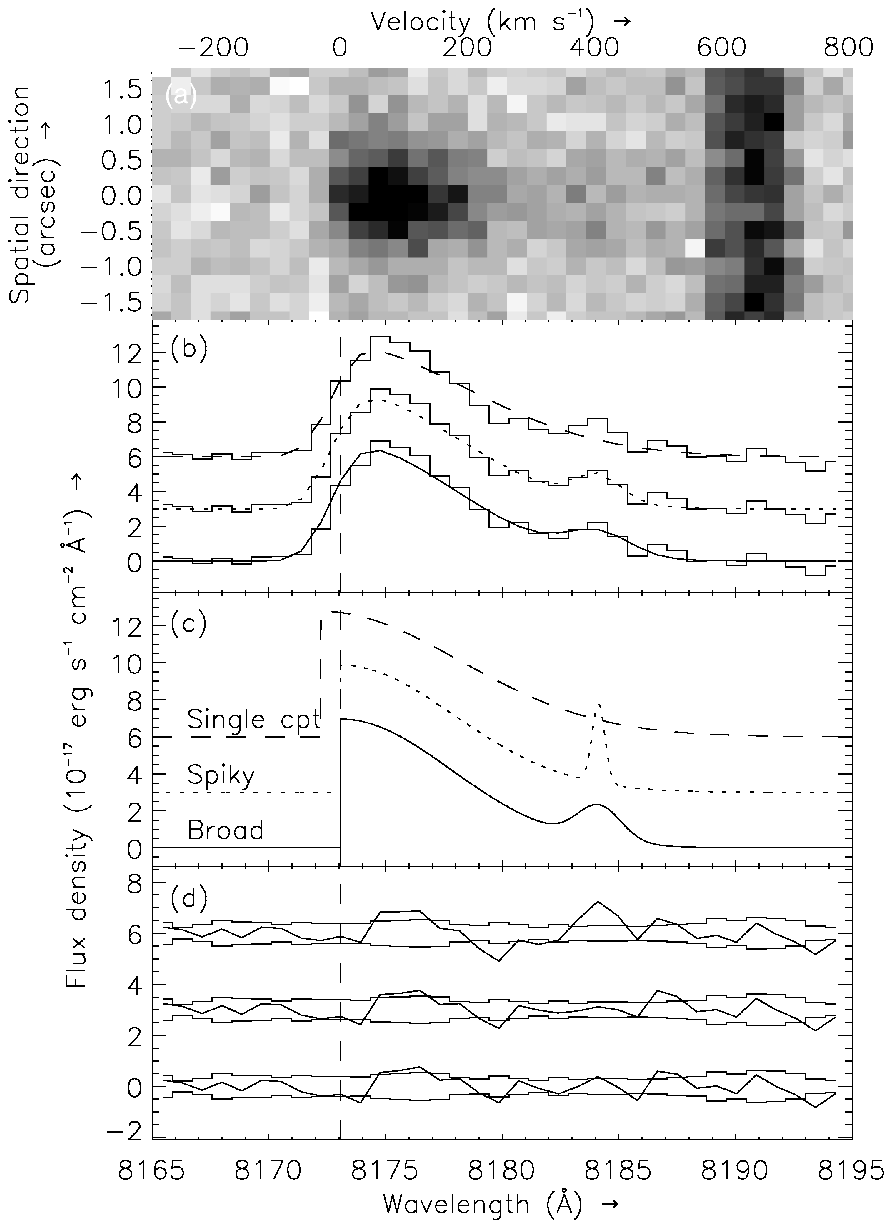,width=0.7\textwidth,trim=110pt 0pt 135pt 0pt,clip}
  \end{center}
  \caption{({\it left}) Thumbnail images of the confirmed \lya\ galaxy
    J114334.98$-$014433.9 in different filters, from top to bottom:
    intermediate band filter \med, narrowband filters \na, \nb, \nc\
    and broadband filters $B$ and $R$. ({\it right}) A fit of the
    \lya\ line with one- and two-component models. ($a$) The two
    dimensional spectrum centered on the \lya\ line. In this
    unrectified spectrum, the night sky emission lines have not been
    removed. The pixel scale is 0\farcs252\,pix$^{-1}$ in the spatial
    direction and 0.86\,\AA\,pix$^{-1}$ in the dispersion direction.
    ($b$) Observed \lya\ line ({\it histograms}) with the three
    best-fitting models. For clarity, the models are offset by
    3\,$\times$\,10$^{-17}$\,\ergsPerAng. The two-component models
    consist of a broad, truncated Gaussian and a narrower redshifted
    Gaussian. The one-component model consists of only a broad,
    truncated Gaussian.  ($c$) Same model line profiles as in ($b$)
    but before convolution with the instrument profile.  ($d$)
    Observed data minus model fit (as plotted in ($b$)) residuals,
    demonstrating a random scatter about the zero flux line.  Also
    shown ({\it histograms}) is the 1$\sigma$-error spectrum from the
    observed data, which includes poissonian noise from both the sky
    and the object. Note that the red peak is not the \ion{N}{v} line.
    At this redshift it would appear around 8334\,\AA.}
  \label{fig:linefit}
\end{figure}

\section{Confirmed \lya\ Emitter at $z=5.721$}
\label{sec:emitter}
One of the candidate \lya\ emitting galaxies (see Figure
\ref{fig:linefit} for the thumbnails) has been confirmed with
follow-up spectroscopy \cite{Westra05}. The object shows a strong
detection in only one of the narrowband images, \nb, and also in the
intermediate band image, \med. In both $B$ and $R$ there is no
detection at the location of the object. In March 2004 the object
J114334.98$-$014433.9 was observed with VLT/FORS2 at medium resolution
(R\,$\sim$\,3600). In the spectrum only one emission line is visible
(see Figure \ref{fig:linefit}). This line is clearly asymmetric, which
is a strong indication for \lya\ \cite{Stern00}. Additionally, the
lack of any other emission lines in our spectrum, such as \hb, \oii,
or \oiii, support that the line is \lya\ and not an emission line at
lower redshift, e.g. \ha\ at $z\sim0.24$. The line is not \Oii,
because the resolution of the spectrum is high enough to resolve the
doublet. Based on these arguments, we conclude that the emission line
is \lya.

From the spectrum we derived a line flux of
5$\times$10$^{-17}$\,\ergscm2, making it one of the brightest \lya\
emitters at this redshift (cf. \cite{Ajiki03,Hu04}). This translates
to a line luminosity of 1.8$\times$10$^{43}$\,\ergs, which is in
accordance with the original survey goals of large volume and bright
survey limit.

We fitted the emission line in a similar fashion as \cite{Hu04} with a
truncated Gaussian (see Figure \ref{fig:linefit}b,c). If we add a
second, narrower and weaker Gaussian to the model, we are able to fit
the second peak in the red wing of the line much better. Fitting in
this way gave us two solutions, both with the second peak at
$\sim$400\,\kms\ redshifted from the truncated Gaussian profile, but
one being broader and less strong than the other. We are unable to
distinguish between the two models given their similar $\chi_\nu^2$.
In general, a second peak in the \lya\ line is a clear signature of an
expanding shell of neutral hydrogen \cite{Ahn03,Dawson02}.

\section{Results of WFILAS}
\label{sec:results}
WFILAS has yielded 7 candidate \lya\ emitting galaxies at $z\sim5.7$
in the range of 1$-$3$\times$10$^{43}$\,\ergs,
corresponding\footnote{To convert between luminosity and star
  formation rate, we use the conversion rate
  SFR(\lya)\,=\,9.1$\times$10$^{-43}$\,L(\lya)\,\Msunyr, from
  \cite{Ajiki03}} to $\sim$10$-$30\,\Msunyr. One of these is the
confirmed \lya\ emitter at $z\sim5.721$ (Section \ref{sec:emitter}).
The remaining candidates await spectral confirmation.

\begin{figure}[!htbp]
  \begin{center}
    \epsfig{figure=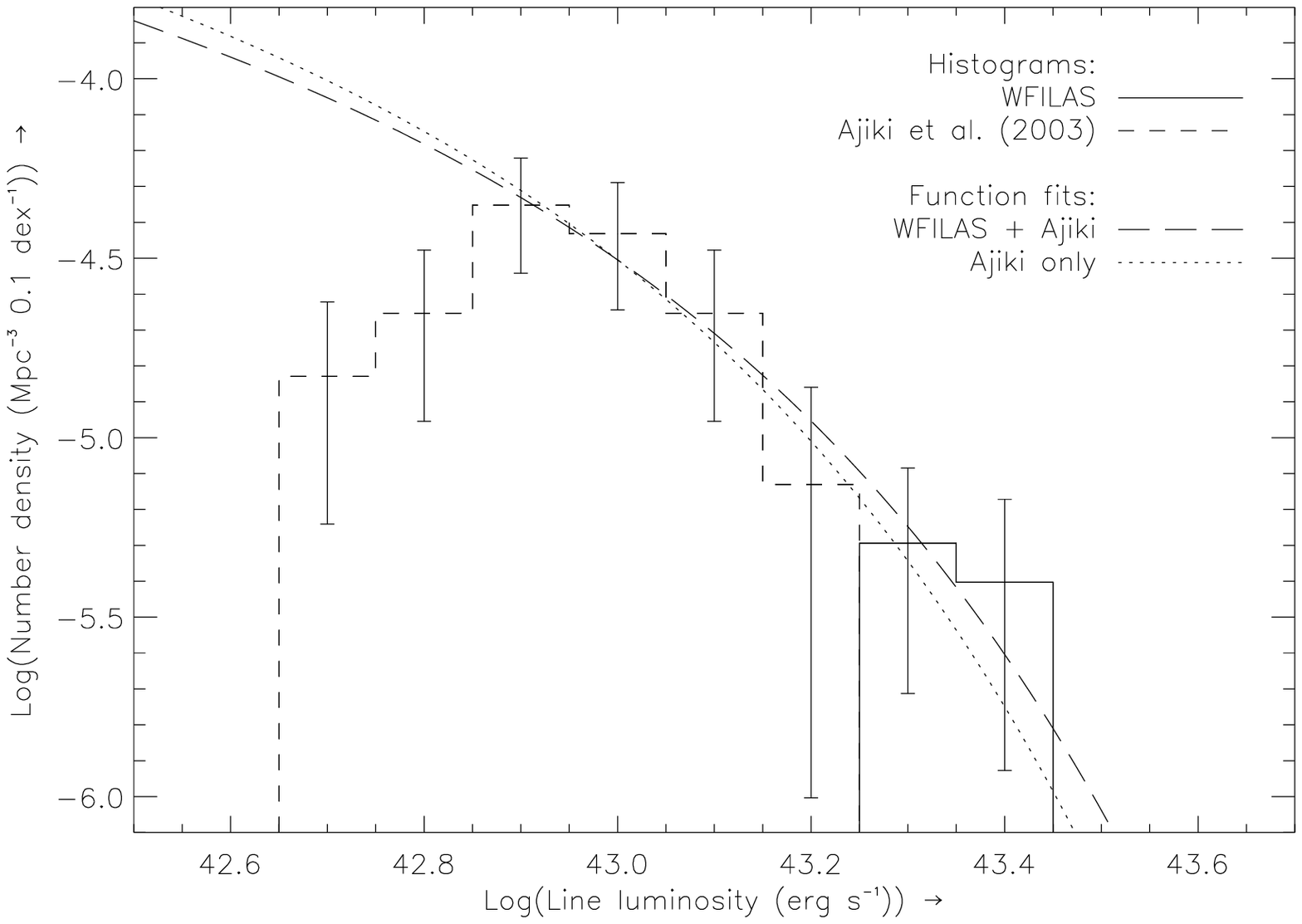,width=0.55\textwidth, trim=15pt 9pt 19pt 20pt, clip}
    \epsfig{figure=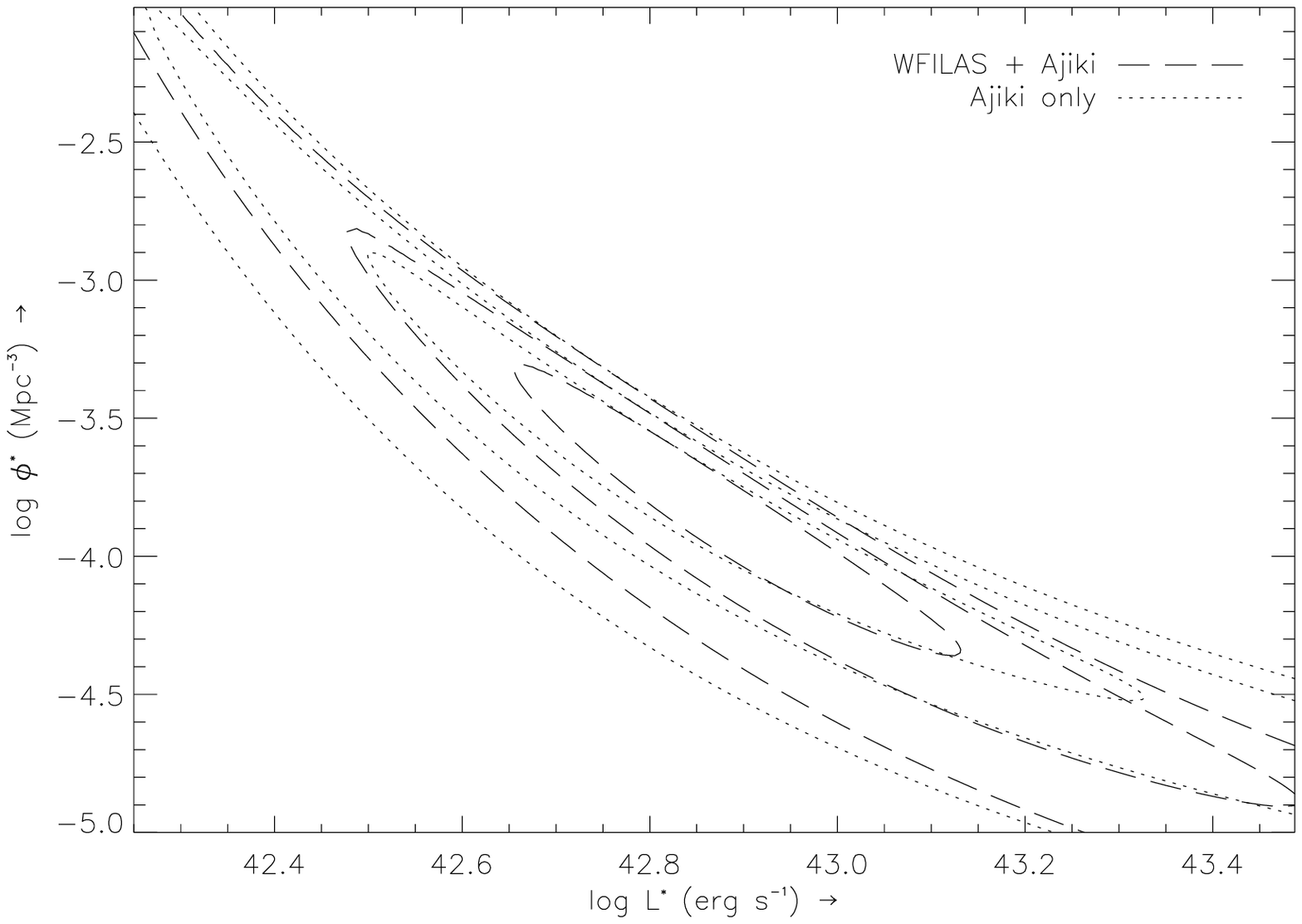,width=0.44\textwidth, trim=15pt 9pt 19pt 20pt, clip}
  \end{center}
  \caption{Line luminosity distribution of the complete sample of
    \lya\ candidates ({\it solid histogram}) together with the
    candidates from \cite{Ajiki03} ({\it dashed histogram}) are
    plotted in the left figure. Both samples are corrected for
    completeness. The errorbars are derived using Poisson statistics.
    Furthermore, two Schechter function fits are indicated: one for
    the combined WFILAS and Ajiki sample ({\it dotted}) and one for
    the Ajiki sample alone ({\it long dashed}).  The right figure
    shows the 68.3\%, 95.4\% and 99.7\% confidence limits for the
    fitting parameters $L^*$ and $\phi^*$. As there is little data at
    faint luminosities, $phi^*$ and $L^*$ are highly correlated.}
  \label{fig:lumfie}
\end{figure}

We have defined two samples: the {\it entire} sample and a {\it
  statistically complete} sample. The entire sample contains all the
candidates. The statistically complete sample contains the candidates
from our four deepest narrowband images. The candidates in the
complete sample are also above a flux limit, which is set at $M_{\rm
  AB}$\,=\,23.38.  Finally, we corrected the sample for completeness.
In Figure \ref{fig:lumfie} we have combined our completeness-corrected
sample with a completeness-corrected sample from Ajiki \cite{Ajiki03}.
Also shown are two Schechter function fits: one for the Ajiki sample
alone, without the lowest two luminosity bins and a fit for the
combined sample of WFILAS and Ajiki (without the lowest two luminosity
bins).  For both fits we assumed a slope of $\alpha$\,=\,-1.53, which
\cite{Ajiki03} adopted from the \ha\ luminosity function at
$z\sim0.24$ from \cite{Fujita03}, since neither sample goes deep
enough to constrain this part of the luminosity function. It can be
seen that the WFILAS sample complements other surveys, which do not
cover a volume as large as that covered by WFILAS.

We note that the number of \lya\ emitting galaxies in the two faintest
bins is considerably less than that predicteed by the fitted
luminosity function. If this is not incompleteness, then the reason
for the drop in the number density of faint sources needs to be
understood. It might be related to the star formation rate. If the
rate is too low, the associated \ion{H}{ii} region will be too small
and most \lya\ photons might be unable to escape the slowly expanding
envelope of neutral hydrogen that surrounds the \ion{H}{ii} region.

\begin{figure}[!htbp]
  \begin{center}
    \epsfig{figure=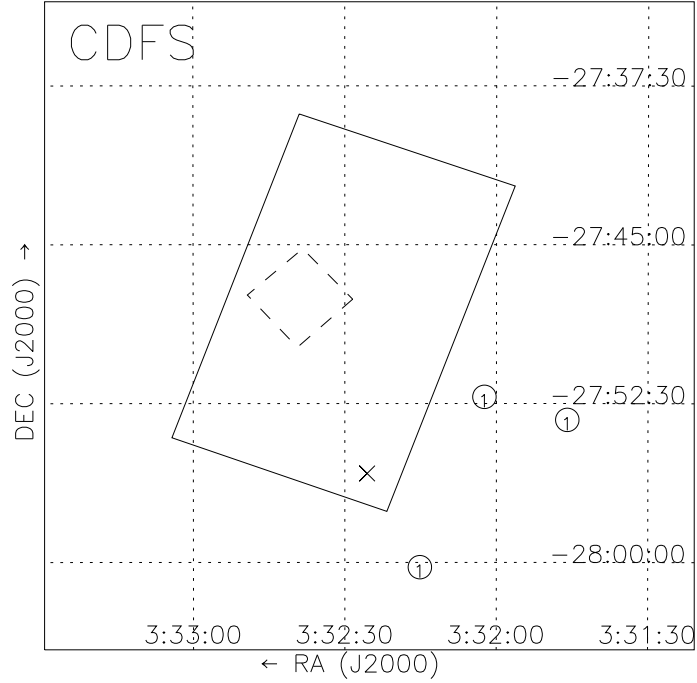,width=0.55\textwidth, clip}
  \end{center}
  \caption{Sky distribution of candidate line emitters in the CDFS
    with North up and East to the left. The designations `1', `2' and
    `3' correspond to \na, \nb\ and \nc\ detected candidates,
    respectively. The gridlines are separated by 7\farcm5. In this
    field the Hubble Ultra Deep Field (HUDF; {\it dashed}) and GOODS-S
    ({\it solid}) are also shown, together with the confirmed $i$-drop
    galaxy at $z=5.78$ of \cite{Bunker03} ({\it cross}). In the CDFS
    there seems to be an overdensity of candidates towards the
    southern part of the field, similar to \cite{Wang05} and recent
    results from the HUDF \cite{Malhotra05}. No candidates were
    detected in the \nb\ and \nc\ filter.}
  \label{fig:overdensity}
\end{figure}

Another result from the WFILAS survey is that in one of the fields we
find an overdensity of sources in the south-eastern part of our
fields, the CDFS (see Figure \ref{fig:overdensity}). To enhance this
result, we have added to our candidates the {\it i$'$}-dropout of
\cite{Bunker03} at $z=5.78$, which is located in the same area of the
field. This overdensity is entirely in agreement with the results of
\cite{Wang05} and the recent discovery of an overdensity at a similar
redshift in the Hubble Ultra Deep Field, which is enclosed in our
field. This result demonstrates the importance of having a wide field
of view not only to find the bright end of the luminosity function,
but also to elucidate variations in large scale structures.

\section{Taurus Tuneable Filter Field Galaxy Survey}
\label{sec:ttffgs}
One of the advantages of WFILAS is that it is also suitable to pick up
low redshift galaxies (see Section \ref{sec:wfilas} and Figure
\ref{fig:examples}). In addition to this, we will use follow-up
spectroscopy to the Taurus Tuneable Filter Field Galaxy Survey
(TTFFGS; \cite{Jones01}). The TTFFGS is a survey designed to find low
redshift emission line galaxies, such as \ha, \hb, \oii\ and \oiii.
For the survey we used the (now decommissioned) Taurus Tuneable Filter
at the Anglo-Australian Telescope. The filter is a Fabry-Perot
interferometer able to create very narrow filter passbands (FWHM
$\sim$20$-$30\AA) with an Airy profile at a user-specified central
wavelength. An order-sorting filter is necessary to remove other
passbands due to multiple interference orders. The TTFFGS scanned
through three OH-airglow windows at 7070\,\AA, 8150\,\AA\ and
9090\,\AA, which effectively creates several narrowband filters next
to each other in each window. For \ha\ these intervals correspond to
$z\sim0.08$, 0.24 and 0.4, respectively. A total of 13 fields have
been observed, each with a diameter of 9\arcmin, giving a total of
\mbox{$\sim$\,825 sq.\,arcmin}. Forty percent of these fields have
been targetted for spectroscopic follow-up with FORS2 on the VLT.
These data are currently being processed.

\acknowledgements{EW wishes to acknowledge the Astronomical Society of
  Australia Travel Grant and the Alex Rodgers Travelling Scholarship
  for financial support.}

\vfill 
\end{document}